

\magnification\magstep1
\parskip=\medskipamount
\hsize=6 truein
\vsize=8.2 truein
\hoffset=.2 truein
\voffset=0.4truein
\baselineskip=14pt
\tolerance=500


\font\titlefont=cmbx12
\font\abstractfont=cmr10 at 10 truept
\font\authorfont=cmcsc10
\font\addressfont=cmsl10 at 10 truept
\font\smallbf=cmbx10 at 10 truept
 4
\font\big=cmr10 scaled \magstep2


\outer\def\beginsection#1\par{\vskip0pt plus.2\vsize\penalty-150
\vskip0pt plus-.2\vsize\vskip1.2truecm\vskip\parskip
\message{#1}\leftline{\bf#1}\nobreak\smallskip\noindent}


\newdimen\itemindent \itemindent=13pt
\def\textindent#1{\parindent=\itemindent\let\par=\resetpar%
\indent\llap{#1\enspace}\ignorespaces}

\let\oldpar=\par
\def\resetpar{\oldpar\parindent=0pt\let\par=\oldpar}

\font\ninerm=cmr9 \font\ninesy=cmsy9
\font\eightrm=cmr8 \font\sixrm=cmr6
\font\eighti=cmmi8 \font\sixi=cmmi6
\font\eightsy=cmsy8 \font\sixsy=cmsy6
\font\eightbf=cmbx8 \font\sixbf=cmbx6
\font\eightit=cmti8
\def\eightpoint{\def\rm{\fam0\eightrm}
  \textfont0=\eightrm \scriptfont0=\sixrm \scriptscriptfont0=\fiverm
  \textfont1=\eighti  \scriptfont1=\sixi  \scriptscriptfont1=\fivei
  \textfont2=\eightsy \scriptfont2=\sixsy \scriptscriptfont2=\fivesy
  \textfont3=\tenex   \scriptfont3=\tenex \scriptscriptfont3=\tenex
  \textfont\itfam=\eightit  \def\it{\fam\itfam\eightit}%
  \textfont\bffam=\eightbf  \scriptfont\bffam=\sixbf
  \scriptscriptfont\bffam=\fivebf  \def\bf{\fam\bffam\eightbf}%
  \normalbaselineskip=9pt
  \setbox\strutbox=\hbox{\vrule height7pt depth2pt width0pt}%
  \let\big=\eightbig \normalbaselines\rm}
\catcode`@=11 %
\def\eightbig#1{{\hbox{$\textfont0=\ninerm\textfont2=\ninesy
  \left#1\vbox to6.5pt{}\right.\n@space$}}}
\def\vfootnote#1{\insert\footins\bgroup\eightpoint
  \interlinepenalty=\interfootnotelinepenalty
  \splittopskip=\ht\strutbox %
  \splitmaxdepth=\dp\strutbox %
  \leftskip=0pt \rightskip=0pt \spaceskip=0pt \xspaceskip=0pt
  \textindent{#1}\footstrut\futurelet\next\fo@t}
\catcode`@=12 %

\def\S{\hbox{{$\Sigma$}}}
\def\G{\hbox{${\cal G}$}}
\def\Q{\hbox{${\cal Q}$}}
\def\Su{\hbox{${\cal S}$}}


\rightline{Freiburg THEP-93/26}
\rightline{gr-qc/9311017}
\bigskip
\centerline{\titlefont
WHAT IS THE GEOMETRY OF SUPERSPACE ? \footnote*{
Contribution to the Proceedings of the conference on Mach's Principle:
``From Newtons Bucket to Quantum Gravity'', held at T\"ubingen,
Germany, July 26-30, 1993.}}

\vskip 1.1 truecm plus .3 truecm minus .2 truecm

\centerline{\authorfont Domenico Giulini}
\vskip 2 truemm
{\baselineskip=12truept
\addressfont
\centerline{Fakult\"at f\"ur Physik,
Universit\"at Freiburg}
\centerline{Hermann-Herder Strasse 3, D-79104 Freiburg, Germany}
}
\vskip 1.5 truecm plus .3 truecm minus .2 truecm

\centerline{\smallbf Abstract}
\vskip 1 truemm
{\baselineskip=12truept
\leftskip=4truepc
\rightskip=4truepc
\parindent=0pt

{\abstractfont
We investigate certain properties of the Wheeler-DeWitt metric
(for constant lapse) in canonical General Realtivity associated
with its non-definite nature.
\par}}

\vskip 1.5truecm

\beginsection{1. Introduction}

As is well known, the dynamics of General Relativity can be
formulated in terms of a constrained
Hamiltonian system, with the configuration space for
pure gravity being given by the space of all Riemannian metrics on a
3-dimensional manifold $\S$ of fixed but arbitrary topology. We call
this space $\Q(\S)$ to indicate its dependence upon the choice of $\S$.
In this Hamiltonian picture, space-time is looked upon as a history of
dynamically evolving geometries on $\S$ represented by a path
$g_{ab}(s)$ in $\Q(\S)$. In the special gauge where the lapse function
$N=1$ and the shift vector $N^a=0$, the vacuum Einstein equations without
cosmological constant decompose into the dynamical part (in units where
$16\pi G/c^4=1$; a dot means differentiation with respect to the parameter
$s$).
$$
{\ddot g}_{ab}+\Gamma_{ab}^{ij\,kl}{\dot g}_{ij}{\dot g}_{kl}=
-2 (R_{ab}-{1\over 4}g_{ab}R)\,,
\eqno{(1)}
$$
and the constraint part
$$\eqalignno{
G^{ab\,cd}{\dot g}_{ab}{\dot g}_{cd}-4\sqrt{g}R & =0
\quad\hbox{Hamiltonian Constraint}              & (2)\cr
G^{ab\,cd}\nabla_b{\dot g}_{cd}                 & =0
\quad\hbox{Momentum Constraint.}                 & (3)\cr}
$$
$R_{ab}$ and $R$ are the Ricci-tensor and Ricci scalar of the metric
$g_{ab}$. $G^{ab\,cd}$ is the DeWitt metric (DeWitt 1967) on the space of
symmetric positive definite matrices (defined below as $G_{\beta}^{ab\,cd}$
for $\beta=1$). The $\Gamma$-symbols in (1) are the Christoffel
symbols for the DeWitt metric. If (2)(3) are satisfied initially
it follows from (1) that they continue to be satisfied throughout the
evolution. (1) and (2) have an obvious geometric interpretation, whereas
(3) says that the velocity must be orthogonal to the orbits of the
diffeomorphism group. This is explained in more detail below.

Due to diffeomorphism invariance, $\Q(\S)$ is endowed with an
action of the Diffeomorphism group $D(\S)$ of $\S$: each point of
$Q(\S)$ is a Riemannian metric on $\S$ which is acted upon by a
diffeomorphism via pull-back. To different metrics which are connected
by a diffeomorphism in such a way are considered to be physically
indistinguishable. Redundancies of this sort are avoided by going to the
quotient $\Su(\S):=\Q(\S)/D(\S)$, called the superspace associated to
$\S$. It represents the space of geometries rather than metrics on $\S$.
Although superspace now faithfully labels physical configurations,
paths in superspace do not faithfully represent space-times. Two
{\it different} paths of geometries may be obtained by ``waving'' $\S$
differently through the {\it same} space-time. But not every path
in $\Su(\S)$ can be obtained by appropriately ``waving'' $\S$
through a given space-time. The former freedom is precisely the freedom
in the choice of the lapse function.

The existence of some geometric structures of superspace is implicit in
many of the investigations into the dynamical structure of General
Relativity. So for example in John Wheeler's view of General Relativity
as Geometrodynamics (Wheeler 1968)
and the associated quantization programme, where superspace serves as domain
for the quantum mechanical state functional. The equations to be satisfied
by this state functional, the Wheeler-DeWitt equations, explicitly
refer to the metric (DeWitt 1967 and Wheeler 1968), just like the
classical equation (1).
Julian Barbour sees the fulfilment of the Machian requirement
on General Relativity in a sucessfull formulation of dynamics solely within
superspace (Barbour 1993). The dynamical principle envisaged is a kind of
geodesic equation with respect to some generalized metric on
superspace (Barbour 1993). All these attempts motivate to have a closer
look at some of the metric structures of superspace.

So we first ask: ``What is the geometry of $Q(\S)$''?
Mathematically there is a variety of possibilities to endow $\Q(\S)$
with a geometry. On the other hand, the laws of General Relativity select
a family of such metrics, one for each choice of the lapse
function $N$. For the particular choice $N=1$ this is displayed
in equations (1)-(3). They define a metric on $\Q(\S)$:
$$
\G(h,k):=\int_{\S}G^{ab\,cd}\,h_{ab}\,k_{cd}\,d^3x\,,
\eqno{(4)}
$$
which we call the Wheeler-DeWitt (WDW) metric.  In this article we
investigate some properties of this particular metric connected with
its indefinite nature.

Note that due to the constraint (3), General Relativity only uses the
WDW metric to calculate inner products on the subspace of tangent
vectors satisfying (3), which requires those vectors to be WDW-orthogonal
to the directions of the diffeomorphisms. We call the diffeomorphism
directions vertical and the WDW-orthogonal directions horizontal.
Due to the indefinite nature of the WDW metric, the horizontal subspace
might also contain vertical directions. When this is not the case, the WDW
metric restricted to the horizontal subspace defines a metric on the quotient
space $\Su(\S)$. But what generally happens is that in different regions
of superspace this quotient-space metric has different signatures.
Such signature changes are precisely signalled by non-trivial intersections
of vertical with horizontal subspaces. To clarify the WDW geometry of
superspace would mean to: 1) characterize
the singular set in $\Q(\S)$ which consists of those points where horizontal
and vertical subspaces intersect non-trivially, and 2) study the restriction
of the WDW-metric to the horizontal subspaces.
Only partial results are known so far.
Note that we do not consider the constraint equation (2) in the same way
as we did with (3). This would select
a non-linear subspace of vectors and thus prevent us from having a
pseudo-Riemannian structure. In this respect we differ from the approach
taken by Barbour (Barbour 1993).

What we wish to show here is that the WDW metric has rather special
properties. This we do by introducing a 1-parameter family of fiducial
metrics of which the WDW metric is one member. The parameter will be called
$\beta$ and the WDW metric is obtained for $\beta=1$.

\beginsection{2. Ultralocal Metrics}

In order to do differential geometry on $\Q(\S)$ we heuristically assume
that $\Q(\S)$ is a differentiable manifold with tangent space
$T_g(Q)$ and cotangent space $T^*_g(Q)$ at the metric $g_{ab}\in Q$
(we shall sometimes drop the reference to $\S$). Elements of $T_g(\Q)$ are
any symmetric covariant tensor field and elements of $T^*_g(\S)$ are any
symmetric contravariant tensor density of weight one on $\S$.
Suppose we want to define
a metric, i.e. a non-degenerate bilinear form in each $T_g(\Q)$. Then up to
an overall constant there is a unique 1-parameter family of ultralocal
metrics (i.e. depending locally on $g_{ab}$ but not on its derivatives)
defined in the following way: take $h,k\in T_g(\Q)$, then
$$\eqalignno{
&\G_{\beta}(h,k):=\int_{\S}G_{\beta}^{ab\,cd} h_{ab} k_{cd}\,d^3x\,,
&(5)\cr
&\hbox{where}\quad
G_{\beta}^{ab\,cd}={\sqrt{g}\over 2}(g^{ac}g^{bd}+g^{ad}g^{bc}-
2\beta g^{ab}g^{cd})\,. &(6)\cr}
$$
The WDW metric, introduced in (4), is just $\G_{1}$.
Given $p,q\in T^*_g(\S)$, the ``inverse'' metric, $\G^{-1}_{\beta}$, is
$$\eqalignno{
&\G^{-1}_{\beta}(p,q):=
\int_{\S} G^{\beta}_{ab\,cd}\, p^{ab}q^{cd}\, d^3x\,, &(7) \cr
&\hbox{where}\quad
G^{\beta}_{ab\,cd}={1\over 2\sqrt{g}}(g_{ac}g_{bd}+g_{ad}g_{bc}-
2\alpha g_{ab}g_{cd}) & (8)\cr
& \hbox{with}\quad \alpha+\beta=3\alpha\beta\,,\quad
\hbox{so that}\quad G_{\beta}^{ab\,nm}G^{\beta}_{cd\,nm}=
{1\over 2}(\delta^a_c\delta^b_d+\delta^a_d\delta^b_c)\,. &(9)\cr}
$$
These are non-degenerate bilinear forms for $\beta\not = 1/3$ (we exclude
$\beta=1/3$), positive definite for $\beta < 1/3$ and of mixed signature
for $\beta> 1/3$ with infinitely many plus as well as minus signs.
Because they are
ultralocal, they arise from metrics on the space $S^+_3$ of symmetric
positive definite matrices -- which is diffeomorphic to the homogeneous
space $GL(3,R)/SO(3)\cong R^6$ -- carrying the metric $G_{\beta}$. One has
$GL(3,R)/SO(3)\cong SL(3,R)/SO(3)\times R^+\cong R^5\times R^+$ and with
respect to this decomposition the metric has a simple warped-product form
$$
G_{\beta}^{ab\,cd}dg_{ab}\otimes dg_{cd}=-\epsilon
d\tau\otimes d\tau+{\tau^2\over c^2}
tr(r^{-1}dr\otimes r^{-1}dr)\,,
\eqno{(10)}
$$
$$
\hbox{with}\quad
c^2=16\vert\beta-1/3\vert,\quad
\tau=cg^{1\over 4},\quad
r_{ab}=g^{-{1\over 3}}g_{ab},\quad
\epsilon=\hbox{sign}(\beta-1/3).
\eqno{(11)}
$$
The matrices $r_{ab}$ are just the coordinates on $SL(3,R)/SO(3)$ and the
trace in (10) is just the left-$SL(3,R)$ invariant metric on this space. This
gives rise to 8 Killing vectors of $G_{\beta}$. An additional homothety is
generated by the multiplicative action of $R^+$ on the $\tau$ coordinate.
Moreover, geodesics in this metric can be explicitly determined
(DeWitt 1967).
If we now regard $Q(\S)$ as a mapping space, i.e. as the space of all
smooth mappings from $\S$ into $S^+_3$, endowed with the metric (5), then,
due to its ultralocal nature, geometric structures like Killing fields,
homotheties and geodesics of the ``target'' metric (10) are inherited by
the full metric (5). For example, dragging the maps $g_{ab}(x)$ along a
Killing flow in
$S^+_3$ produces a Killing flow in $\Q(\S)$. In this way, some geometry of the
infinite dimensional $\Q(\S)$ can be studied by looking at the 6-dimensional
$S^+_3$.

Note also that expression (5) is invariant under diffeomorphisms of $\S$.
An infinitesimal diffeomorphism is represented by a vector field $\xi$
on $\S$ and gives rise to a vector field $X^{\xi}$ on $\Q(\S)$:
$$
X^{\xi}_{ab}=\nabla_a\xi_b+\nabla_b\xi_a\,,
\eqno{(12)}
$$
which is a Killing field of the metric (5). The totalitiy of vectors of
the form (12) at $g\in\Q(\S)$ span what we call the vertical vector space
$V_g\subset T_g(\Q)$. With respect to $\G_{\beta}$ we can define the
orthogonal complement to $V_g$ which we call the horizontal vector space
$H^{\beta}_g\subset T_g(\Q)$. From (5)(6) and (12) we have
$$
k_{ab}\in H^{\beta}_g\,\Leftrightarrow\,
\nabla^a\left(k_{ab}-\beta g_{ab}k_c^c\right)=0\,.
\eqno{(13)}
$$
Under the isometric action of $D(\S)$ on $\Q(\S)$ horizontal
spaces are clearly mapped into horizontal spaces.

If we set $\beta=0$, the metric (5) is positive definite such that
orthogonality also implies transversality, i.e. $V_g\cap H^0_g=\{0\}$.
It is in fact true that the tangent space splits into the direct sum
of closed orthogonal subspaces: $T_g(\S)=V_g\oplus H^0_g$. This allows
to define a Riemannian geometry on the quotient space $\Su(\S)$ by
identifying its tangent spaces with the horizontal spaces in $T(\Q)$
(Ebin 1970).
(Here we pretend $\Su(\S)$ being a genuine manifold).
This works for all $\beta<1/3$. We are, however, interested in
the range $1/3\leq\beta\leq 1$ with special attention paid to
the transition $\beta<1$ to $\beta=1$.

For $\beta>1/3$ the metric (5) is not definite anymore such that
generally $V_g\cap H^{\beta}_g\not =\{0\}$ for such $\beta$.
A simple example is the following: Take as $\S$ a 3-manifold
that carries a flat metric $g$. In
$T_g(\S)$ consider the infinite dimensional vector subspace given by all
vectors of the form $k_{ab}=\nabla_a\nabla_b\phi$, where $\phi$ is a
smooth function on $\S$. These vectors satisfy (13) for $\beta =1$ and are
therefore in $H_g^1$. But they are also of the form (12), with
$2\xi_a=\nabla_a\phi$, and hence in $V_g$. Moreover, suppose the metric is
only flat in an open subset $U\subset\S$. Then we can repeat the argument
but this time only using functions $\phi$ with compact support inside $U$.
Again, these give rise to an
infinite intersection $V_g\cap H_g^{1}$ for each such partially flat
metric $g$.
Clearly, vectors in $H^{\beta}_g\cap V_g$ are necessarily of zero
$\G_\beta$-norm.

\beginsection{3. Some Observations Concerning the WDW Metric}

It follows from (12) and (13) that a vertical vector $X^{\xi}$ is
horizontal, if and only if
$$
D_{\beta}\xi_a:=-\nabla^b(\nabla_b\xi_a-\nabla_a\xi_b)-
2(1-\beta)\nabla_a\nabla^b\xi_b-2R^b_a\xi_b=0\,,
\eqno{(14)}
$$
where $R^b_a$ denote the mixed components of the Ricci-tensor. Killing
vectors, if existent, are obvious solutions but these do not interest us
since they correspond to zero $X^{\xi}$. For $0\leq\beta<1/3$ these are the
only solutions. This implies that for $\beta>1/3$ any non-Killing solution
must have non-zero divergence, since for zero divergence
fields the $\beta$ dependence in (14) drops out.
A more elegant way to write $D_\beta$ is, using the exterior derivative $d$,
its adjoint $\delta$ (given by minus the divergence on the first index) and
writing $Ric$ for the map induced by $R^b_a$:
$$
D_\beta=\delta d + 2(1-\beta)d\delta-2Ric  \,,
\eqno{(15)}
$$
which also displays its formal self-adjointness.
The $\G_{\beta}$-norm of $X^{\xi}$ is given by
$$
\G_{\beta}(X^{\xi},X^{\xi})=2\int_{\S}\xi^aD_{\beta}\xi_a\,d^3x\,.
\eqno{(16)}
$$
For $\beta\leq 1$ and $Ric<0$ (i.e. strictly negative eigenvalues) this
operator is manifestly positive and
$\G_{\beta}$ restricted to $V_g$ is thus positive definite. In particular,
we have $V_g\cap H_g^{\beta}=\{0\}$ for all $g$ such that $Ric<0$ and
$\beta \leq 1$. Since it is known that any 3-manifold $\S$ admits such
Ricci-negative metrics (Gao and Yau 1986),
this tells us that in every superspace there are
open regions (the Ricci-negative geometries) with well defined WDW metric,
given by the restriction of $\G_{1}$ to $H_g^{1}$, whose signature has
infinitely many plus and minus signs.

For a flat metric $g$ and values $\beta<1$, $D_\beta$ is non-negative
with kernel given by the covariantly constant $\xi$. Indeed, from (11)
it follows that $\xi$ is curl- and divergence-free on a flat manifold,
hence covariantly constant. But this also means that $\xi$ is Killing
and therefore $X^{\xi}$ zero. So for $g$ flat we have
$V_g\cap H_g^{\beta}=\{0\}$ for $\beta<1$. On the other hand, for
$\beta=1$ and $g$ flat, we can only infer from (15)
that $\xi$ must be closed, hence exact or harmonic.
But harmonicity implies Killing so that all horizontal $X^{\xi}$
are given by the expressions anticipated in the previous section.
As stated there, we can localize the construction and obtain an infinite
subspace in the intersection $V_g\cap H_g^1$ for metrics $g$ which contain
a flat region $U\subset\S$. Clearly, any manifold admits such metrics.
In particular, this tells us that in every superspace there are regions
where no WDW metric is defined.

It is more difficult to obtain general results for metrics which are
neither Ricci-negative nor flat. For the very special class of
non-flat Einstein $\hbox{metrics}^1$ it is at least easy to see that
for $\beta=1$
$H_g^1\cap V_g$ is zero. Indeed, for $R_{ab}=\lambda g_{ab}$, where
$\lambda\in R-\{0\}$, (15) implies $0=\delta D_1\xi=2\lambda\delta\xi$,
so that $\xi$ must be divergence free and hence $X^{\xi}$ zero. So
there exists a WDW metric for non-flat Einstein geometries in $\Su(\S)$,
given by the restriction of $\G_1$ to $H_g^1$. For the study of such
metrics it is instructive to look at a particular example in detail
to which we now turn.

As non-flat Einstein metric we take the standard round metric on the
three-sphere with some unspecified radius. Here $Ric>0$ and not much
can be directly read off (15) for general $\beta$. But taking elements
of $T_g(\Q)$ as first order perturbations of $g$, and expanding them in
terms of the well known complete set of tensor harmonics
(Gerlach and Sengupta 1978) one can establish the following scenario:
For $1/3<\beta <1$ the number of negative directions (i.e. the number
of linearly independent vectors of negative $\G_{\beta}$-norm)
is finite in $V_g$ and infinite in $H_g^{\beta}$.
For the discrete values $\beta =\beta_n$, where
$$
\beta_n:={n^2-3\over n^2-1},\quad n\in\{3,4,5,\dots\}\,,
\eqno{(17)}
$$
the intersection $V_g\cap H_g^{\beta}$ is non-trivial and of some finite
dimension $d_n>0$. At other values of $\beta$ it is zero. What turns out
to happen is
that when $\beta$ passes the value $\beta_n$ from below, $d_n$ of the
negative directions change from $H_g^{\beta}$ to $V_g$. Since the $\beta_n$
accumulate at $1$, this happens infinitely often as we turn up $\beta$ to
$1$. At $\beta=1$ only a single negative direction has remained in $H_g^{1}$
and infinitely many are now in $V_g$. The intersection $V_g\cap H_g^{1}$ is
in fact zero, in accordance with the more general argument given above.
$\G_{1}$ restricted to $H_g^{1}$
is of Lorentzian signature $(-,+,+,+,\dots)$. This is directly related to
the statement made in quantum cosmology, that the Wheeler-DeWitt
$\hbox{equation}^2$ (for constant lapse) for perturbations
around the three-spehere is hyperbolic
(Halliwell and Hawking 1985). It follows from our
considerations that this can at best be locally valid
since the metric for constant lapse necessarily suffers from signature
$\hbox{changes}^3$. Note also how delicately the signature structure
of $\G_{\beta}$ restricted to $H_g^{\beta}$ depends on whether
$\beta <1$ or $\beta =1$.

There are other interesting differences between $\beta<1$ and
$\beta=1$. Quite striking is the existence of an infinite dimensional
intersection  $H_g^1\cap V_g$ for flat $g$. This means that $D_1$ cannot
be an elliptic operator since these have finite dimensional kernels. And,
in fact, calculating the the principal symbol for $D_{\beta}$ from (14),
we obtain
$$
\sigma_{\beta}(\zeta)^a_b= \Vert\zeta\Vert^2\left(
\delta^a_b+(1-2\beta){\zeta^a\zeta_b\over\Vert\zeta\Vert^2}
\right)\,.
\eqno{(18)}
$$
This matrix is positive definite for $\beta<1$, invertible but not
positive definite for $\beta>1$ and singular positive semi-definite
for $\beta=1$. Expressed in standard terminology this says that the
operator $D_{\beta}$ is strongly elliptic in the first, elliptic
but not strongly elliptic in the second, and degenerate elliptic but
not elliptic in the third case.
This relates to the problem of how one would actually calculate the
metric on superspace at the regular points. Throughout we said that
it would be obtained by restricting the metric $\G_{\beta}$ to the
horizontal spaces $H_g^{\beta}$. But this means that we have to
explicitly calculate the projection $T_g(\Q)\rightarrow H_g^{\beta}$.
A general tangent vector
$k_{ab}\in T_g(\Q)$ is projected by adding a vertical vector $X^{\xi}$
so that the sum is horizontal, i.e. satisfies (13). This is equivalent
to solving
$$
D_{\beta}\xi_b=\nabla^a(k_{ab}-\beta g_{ab}k_c^c)
\eqno{(19)}
$$
as equation for $\xi$ and given right hand side. Uniqueness for $X^{\xi}$
is given at regular geometries, i.e. those for which the kernel of
$D_{\beta}$ consists of Killing vectors only. Since the right hand
side is orthogonal to Killing vectors, ellipticity (for $\beta <1$)
guarantees existence for any $k_{ab}$. It is not clear to us at this
moment whether the failure of ellipticity for $\beta =1$ can in fact
imply any problem. For example, in
the special cases where $g_{ab}$ is an Einstein metric, we can Hodge
decompose $\xi$ and the right hand side of (19) into exact, co-exact and
harmonic forms. The Einstein condition then prevents the Ricci-term in
$D_{1}$ to couple these components, so that (19) decomposes into 3 decoupled
equations for the Hodge modes, two purely algebraic ones and an elliptic
partial differential equation for the co-exact mode. In this case we can
thus show existence by restricting to appropriate subspaces.

\vskip1.5truecm

\noindent
{\it Acknowledgements.} After the talk Karel Kucha\v r pointed
out that observations similar to some of ours were made by
by Friedman and Higuchi (Friedman and Higuchi 1990, appendix).

\beginsection{\authorfont Notes}

${}^1$ In three dimensions an Einstein metric implies constant
sectional curvature so that $\S$ is a space form. But not only
is the topology of $\S$ severly restricted (e.g. its second homotopy
group must be trivial). If $\S$ allows for Einstein metrics,
they only form a finite dimensional subspace in superspace which
is in fact of dimension one if the Einstein constant is non-zero.
In these cases the only deformations are the constant rescalings
of the metric. In this sense Einstein metrics are very special.

${}^2$ There is one Wheeler-DeWitt equation for each smearing function.
If written without smearing functions (as a distribution), the
Wheeler-DeWitt equations for pure gravity looks like an infinite number of
six-dimensional Klein-Gordon equations, one per point $x\in\S$ for
the six components $\{g_{ab}(x)\}$. If added together with a smearing
function, the resulting equation is clearly ultrahyperbolic. Only if
the directions of differentiation are restricted to lie in a horizontal
subspace, or even further, like suggested by Hawking
(Hawking 1983, chapter 5), one may be able to eliminate all but one of
the negative directions. This particular Wheeler-DeWitt equation may then
said to be locally hyperbolic on superspace.

${}^3$ In applications, the Wheeler-DeWitt equations have only been
studied in neighbourhoods of highly symmetric metrics like the one
on the three sphere considered here. It would be interesting to know
how ``far'' from such a point one has to go in order to encounter singular
regions and signature change. The regions $Ric<0$ do not seem ``close'',
and the reason why the Wheeler-DeWitt equations have not been studied in
neighbourhoods of those metrics seems to be the fact that $Ric<0$ metrics
do not allow for any metrics with symmetries.

\beginsection{\authorfont References}

\noindent
Barbour, Julian B. (1993). See contribution to this Volume.

\noindent
DeWitt, Bryce S. (1967). ``Quantum Theory Gravity I. The Canonical
Theory.'' {\it Physical Review} 160: 1113-1148.

\noindent
Ebin, David G. (1970). ``The Manifold of Riemannian Metrics''.
{\it Proceedings of the American Mathematical Society, Symposia
in Pure Mathematics, Global Analysis} 15: 11-40.

\noindent
Friedman, John L. and Higuchi, Atsushi (1990). ``Symmetry and Time on
the Superspace of Asymptotically Flat Geometries''. {\it Physical
Review D} 41: 2479-2486.

\noindent
Gerlach, Ulrich H. and Sengupta, Uday K. (1978). ``Homogeneous
Collapsing Star: Tensor and Vector Harmonics for Matter and Field
Asymmetries''. {\it Physical Review D} 18: 1773-1784.

\noindent
Gao, Zhiyong L., and Yau, Shing-Tung (1986). ``The Existence of
Ricci Negatively Curved Metrics on Three Manifolds''.
{\it Inventiones Mathematicae} 85: 637-652.

\noindent
Halliwell, Jonathan, J. and Hawking, Stephen, W. (1985). ``The Origin
of Structure in the Universe''. {\it Physical Review D} 31: 1777-1791.

\noindent
Hawking, Stephen W. (1984). ``The Quantum State of the Universe''.
{\it Nuclear Physics B} 239: 257-276.

\noindent
Wheeler, John A. (1968). ``Superspace and the Nature of Quantum
Geometrodynamics.'' In {\it Battelle Rencontres}, 1967 Lectures in
Mathematics and Physics. Cecile M. DeWitt and John A. Wheeler, ed.
New York and Amsterdam: W.A. Benjamin Inc., pp. 242-307.

\end